# Directional-dependent thermally activated motion of vortex bundles and theory of anomalous Hall effect in type-II conventional and high-Tc superconductors


Wei Yeu Chen [+] and Ming Ju Chou [*]

Department of Physics, Tamkang University, Tamsui 25137, Taiwan

( 10 November, 2007)


## Abstract


The anomalous Hall effect for type-II conventional and high-$T_C$ superconductors is studied based upon the theory of thermally activated motion of vortex bundles jumping over the directional-dependent energy barrier. It is shown that the Hall anomaly is universal for type-II conventional and high-$T_C$ superconductors as well as for superconducting bulk materials and thin films, provided certain conditions are satisfied. We find that the directional-dependent potential barrier of the vortex bundles renormalizes the Hall and longitudinal resistivities, and Hall anomaly for superconductors is induced by the competition between the Magnus force and the random collective pinning force of the vortex bundle. We also find that the domain of anomalous Hall effect includes two regions: the region of thermally activated motion of the small vortex bundles and that of the large vortex bundles separated by the contour of the quasiorder-disorder first-order phase transition, or the peak effect of the vortex system. The Hall and longitudinal resistivities as functions of temperature as well as applied magnetic field have been calculated for type-II superconducting films and bulk materials. The conditions for occurring the double sign reversal or reentry phenomenon is also investigated. All the results are in agreement with the experiments.





[+] Corresponding author. Tel.: 886-2-28728682; fax: 886-2-28728682

E-mail: wychen@mail.tku.edu.tw (W.Y. Chen)

[*] Present address: Department of Physics, National Taiwan Normal University, Taipei, Taiwan 116




# I. INTRODUCTION

The vortex dynamics in type-II superconductors, discovered by Abrikosov[1], has been focused tremendously efforts[2–36], especially the most confusing and unusual behavior of the Hall resistivity in many high-temperature and in some conventional type-II superconductors[2,3,13–25], since the Hall effect was first measured by Niessen and Staas[2] in 1965, later the Hall anomaly was observed by Van Beelen et al[3]. Experimental data show that when the temperature is slightly below the critical temperature, the Hall resistivity in many type-II superconductors changes its sign as the temperature or applied magnetic field decreases. In some occasions, the observed Hall resistivity exhibits the double sign reversal, or the reentry phenomenon. Various theories, trying to explain this anomaly, have been proposed, such as the large thermomagnetic model[16], bound vortex-antivortex pair[17], flux flow[18], opposing drift of quasiparticles[19], and many others[4]. However, up to the present time, there is no satisfactory explanation for this sign reversal of the Hall resistivity. The origin of this Hall anomaly still remains unsolved.

It is well known that the quenched disorder destroys the long-range order of the flux line lattice, after which only short-range order, the vortex bundle, prevails[5,6]. In this paper, we develop a self-consistent theory for the thermally activated motion of



the vortex bundles, under the steady-state condition, jumping over the directional-dependent potential barrier generated by the Magnus force, the random collective pinning force, and the strong pinning force inside the vortex bundle. The directional-dependent energy barrier means that the energy barrier is different when the direction of thermally activated motion is different. Based on this theory, it is shown that the Hall anomaly is universal for type-II conventional and high-$T_C$ superconductors as well as for superconducting bulk materials and thin films, provided certain conditions are satisfied. We find that the directional-dependent potential barrier of the vortex bundles renormalizes the Hall and longitudinal resistivities, and the anomalous Hall effect is induced by the competition between the Magnus force and the random collective pinning force of the vortex bundle. We also find out that the domain of Hall anomaly includes two regions, the region of thermally activated motion of the small vortex bundles and that of the large vortex bundles separated, by the contour of the quasiorder-disorder first-order phase transition or the peak effect [6] of the vortex system. The region belongs to the thermally activated motion of large vortex bundles for low temperature and applied magnetic field . In this region, the transverse size of the short-range order is approximately $10^{-6}$ m. However, if the applied magnetic field increases for a fixed temperature or the temperature increases for a fixed magnetic field, a quasiorder-disorder first-order phase transition between the



short-range order and disorder in the vortex system, or the peak effect eventually

occurs [6] due to the enormous increase in the dislocations inside the short-range

domains. In this case, the vortex lines become a disordered amorphous vortex system.

However, they are not individual single quantized vortex lines, the vortex lines still

bounded close together to form small vortex bundles of the dimension $R \cong 10^{-8}\, m$, the

region then crosses over to that of the thermally activated motion of small vortex

bundles.

Under the framework of our theory, the Hall and longitudinal resistivities as

functions of temperature as well as applied magnetic field are calculated for type-II

superconducting films and bulk materials. All the results are in agreement with

the experiments. It is worthwhile pointing out that in this paper we concentrate our

efforts on the study of anomalous Hall effect for type-II superconductors only, not

the anomalous Hall effect or spin Hall effect in semiconductors and simple metals.

The rest of this paper is organized as follows. The thermally activated motion of

vortex bundles jumping over the directional-dependent potential barrier from one

equilibrium position to another is developed and the theory of anomalous Hall effect

is presented in Sec. II. Within this theory, the anomalous Hall effect for type-II

superconducting bulk materials is studied in Sec III. In Sec. IV, the corresponding

Hall anomaly for superconducting films is investigated. In Sec. V, the conditions for



occurring of the double sign reversal or reentry phenomenon is studied. Finally, a concluding remarks are conveyed in Sec. VI.

## II. THERMALLY ACTIVATED MOTION OF VORTEX BUNDLES AND THEORY OF ANOMALOUS HALL EFFECT

We shall develop a self-consistent theory for thermally activated motion of vortex bundles, under the steady-state condition, jumping over the directional-dependent potential barrier from one equilibrium position to another. By applying the random walk theorem, we calculate the coherent frequency $v_c$ for the vortex bundles, and evaluate the root-mean-square value of the angle between the random collective pinning force of the vortex bundle and the positive $y$-direction with current in the x-direction. Finally, the Hall and longitudinal resistivities are obtained.

### A. CALCULATION OF COHERENT FREQUENCY OF VORTEX BUNDLES BY RANDOM WALK THEOREM

From the theory of forced oscillations, it is understood that the response function of the vortex line oscillates inside the potential barrier due to thermal agitation. By identifying the oscillation energy of the vortex line inside the potential barrier with the thermal energy, the thermal oscillation frequency of the individual vortex inside the potential barrier can be expressed as

$$v = \bar{v}\sqrt{T} \quad , \tag{1}$$

where the proportional constant $\bar{v}$ is given by $\bar{v} = \dfrac{1}{\pi A}\sqrt{\dfrac{k_B}{2m}}$, $A$ is the average



amplitude of the oscillation, $k_B$ is the Boltzmann constant and $m$ is the mass of the vortex line [20]. It is worthwhile to point out that the viscous damping of the vortex line is included implicitly in the average amplitude $A$. However, the oscillations of vortex lines inside the vortex bundle are not coherent, namely, their oscillations are at random. To obtain the coherent oscillation frequency $v_C$ of the vortex bundle as a whole, by applying the random walk's theorem, the frequency $v$ in equation (1) must be divided by the square root of N, the number of vortices inside the vortex bundle,

$$v_C = \frac{v}{\sqrt{N}} = \frac{\overline{v}\sqrt{T}\sqrt{\Phi_0}}{R\sqrt{\pi B}} \quad , \tag{2}$$

where $\Phi_0$ is the unit flux, $R$ is the transverse size of the vortex bundle, $B$ is the value of the applied magnetic field. The essential property can be comprehended by considering the problem of random walk. Let $l_0$ be the length of each individual step, for a walk with N steps, if the walk were coherently in the same direction, the total length of N steps would be $L = N l_0$; however, these steps are not coherently in the same direction, they are at random. With the aim of receiving the length of the random walk with N steps, the above expression must be divided by the square-root of N

$$L = \frac{N l_0}{\sqrt{N}} = \sqrt{N}\, l_0 \quad , \tag{3}$$

this is the desired answer for the random walk with N steps.



## B. ROOT-MEAN-SQUARE ANGLE BETWEEN VORTEX BUNDLES AND Y-AXIS

Let us consider the case for p-type superconductors with current flowing in the positive x-direction and the applied magnetic field in the positive z-direction. If we assume that the mean angle between the random collective pinning force of a vortex line inside the vortex bundle and the positive y-direction measured in counterclockwise sense is $\theta$, this temperature-and field-dependent $\theta$ can be obtained as follows: Since $\theta$ is small, we can approximately write $\theta \cong \frac{|\vec{f}_{el}|}{|\vec{f}_L|}$, where $|\vec{f}_{el}|$ and $|\vec{f}_L|$ are the magnitudes of the elastic force and the Lorentz force of the vortex line. Taking into account the fact that the compression modulus $C_{11}$ is much larger than the shear modulus $C_{66}$ [6], owing to the thermal fluctuations, the magnitude of the displacement vector $|\vec{S}_f(\vec{r})|$ of the vortex line inside the vortex bundle as well as its corresponding magnitude of the elastic force $|\vec{f}_{el}|$ is proportional to $\sqrt{k_B/C_{66}}$ or $(1/\sqrt{B})\sqrt{T/T_C - T}$ [4,6,28], $T_C$ is the critical temperature of the superconductors. The temperature-and field-dependent $\theta$ can now be expressed as

$$\theta(T,B) = \bar{\alpha}' \frac{1}{\sqrt{B}} \sqrt{\frac{T}{T_C - T}} , \qquad (4)$$

$\bar{\alpha}'$ is a proportional constant. The mean angle $\Theta(T,B)$ between the random collective pinning force of vortex bundle and the positive y-direction measures in



counterclockwise sense, by the theory of random walk, can be written as

$$\Theta(T,B) = \sqrt{N}\,\theta(T,B) = \overline{\alpha}\sqrt{\frac{T}{T_C - T}}, \tag{5}$$

with $\overline{\alpha} = \overline{\alpha}' R\sqrt{\pi/\Phi_0}$. The root-mean-square value of the angle between the random collective pinning force for the vortex bundle and the positive y-direction in counterclockwise sense can now be obtained as

$$\Psi = [\int_{-\frac{\pi}{2}}^{\frac{\pi}{2}} \phi^2 \exp(\frac{-\phi^2}{\Theta^2(T,B)})\,d\phi / \int_{-\frac{\pi}{2}}^{\frac{\pi}{2}} \exp(\frac{-\phi^2}{\Theta^2(T,B)})\,d\phi]^{\frac{1}{2}}. \tag{6}$$

Keeping in mind the fact that $\Theta(T,B)$ is usually very small in our theory, we obtain

$$\Psi \cong \Theta(T,B) = \overline{\alpha}\sqrt{\frac{T}{T_C - T}}. \tag{7}$$

### C. CALCULATION OF HALL AND LONGITUDINAL RESISTIVITIES

To proceed, let us calculate the directional-dependent energy barrier of the vortex bundles formed by the Magnus force, the random collective pinning force, and the strong pinning force inside the vortex bundle for the magnetic field in the z-direction $\vec{B} = B\vec{e}_z$, and the transport current in the x-direction $\vec{J} = J\vec{e}_x$. Considering the case where the magnitude of Lorentz force $JB$ is slightly greater than the magnitude of the random collective pinning force, after some algebra, the directional-dependent energy barrier of the vortex bundles both in the positive and negative x-direction as well as the positive and negative y-direction are obtained



respectively as

$$U + \bar{V}R(JB\frac{v_{by}}{v_T} - <F_{p_x}>_R) \quad , \tag{8}$$

$$U - \bar{V}R(JB\frac{v_{by}}{v_T} - <F_{p_x}>_R) \quad , \tag{9}$$

$$U + \bar{V}R(JB - JB\frac{v_{bx}}{v_T} - <F_{p_y}>_R) \quad , \tag{10}$$

$$U - \bar{V}R(JB - JB\frac{v_{bx}}{v_T} - <F_{p_y}>_R) \quad , \tag{11}$$

where the potential barrier $U$ is generated by the strong pinning force due to the randomly distributed strong pinning sites inside the vortex bundle, $\vec{v}_b$ ($\vec{v}_T$) is the velocity of the vortex bundle (super current), $\bar{V}$ is the volume of the vortex bundle, once again, $R$ represents the transverse size of the vortex bundle, the range of $U$ is assumed to be the order of $R$, and $<\vec{F}_p>_R$ stands for the random average of the random collective pinning force per unit volume. The self-consistent equations for the velocity of the thermally activated motion of the vortex bundles jumping over the directional-dependent energy barrier are therefore obtained in components as

$$v_{bx} = v_C R \{\exp[\frac{-1}{k_B T}(U + \bar{V}R(JB\frac{v_{by}}{v_T} - <F_{p_x}>_R))]$$
$$- \exp[\frac{-1}{k_B T}(U - \bar{V}R(JB\frac{v_{by}}{v_T} - <F_{p_x}>_R))]\} \quad , \tag{12}$$

and

$$v_{by} = v_C R \{\exp[\frac{-1}{k_B T}(U + \bar{V}R(JB - JB\frac{v_{bx}}{v_T} - <F_{p_y}>_R))]$$
$$- \exp[\frac{-1}{k_B T}(U - \bar{V}R(JB - JB\frac{v_{bx}}{v_T} - <F_{p_y}>_R))]\} \quad , \tag{13}$$

where $v_C$ is the coherent oscillation frequency of the vortex bundle jumping over the



directional-dependent potential barrier from one equilibrium position to another.

Taking into account the fact that $\frac{v_{bx}}{v_T} \ll 1$, Eqs. (12) and (13) can be approximately

rewritten as

$$v_{bx} = (\frac{\overline{V}}{R}\sqrt{\frac{T\Phi_0}{\pi B}})\, R\, \exp(\frac{-U}{k_B T})\{\exp[\frac{-\overline{V}\,R}{k_B T}(\frac{-JB|v_{by}|}{v_T} + |<F_p>_R|\sin\Psi)]$$

$$-\exp[\frac{+\overline{V}\,R}{k_B T}(\frac{-JB|v_{by}|}{v_T} + |<F_p>_R|\sin\Psi)]\} \quad , \tag{14}$$

$$v_{by} = (\frac{\overline{V}}{R}\sqrt{\frac{T\Phi_0}{\pi B}})\, R\, \exp(\frac{-U}{k_B T})\{\exp[\frac{-\overline{V}\,R}{k_B T}(JB - |<F_p>_R|\cos\Psi)]$$

$$-\exp[\frac{+\overline{V}\,R}{k_B T}(JB - |<F_p>_R|\cos\Psi)]\} \quad , \tag{15}$$

where $\Psi$ is the root-mean-square value of the angle between the random collective

pinning force of the vortex bundles and the positive $y$-direction measured in the

counterclockwise sense. By considering the identities $\vec{E} = -\vec{v}_b \times \vec{B}$, $\rho_{xx} = \frac{E_x}{J}$,

$\rho_{xy} = \frac{E_y}{J}$, together with Eq. (7), and keeping in mind that $\Psi$ is usually very small,

the longitudinal and Hall resistivities, can now be written respectively as follows,

$$\rho_{xx} = \frac{\overline{V}\sqrt{BT\Phi_0}}{J\sqrt{\pi}}\exp(\frac{-U}{k_B T})\{\exp[\frac{\overline{V}R}{k_B T}(JB - (\frac{\beta^C(T,B)}{\overline{V}})^{\frac{1}{2}})] - \exp[\frac{-\overline{V}R}{k_B T}(JB - (\frac{\beta^C(T,B)}{\overline{V}})^{\frac{1}{2}})]\}, \tag{16}$$

$$\rho_{xy} = \frac{-\overline{V}\sqrt{BT\Phi_0}}{J\sqrt{\pi}}\exp(\frac{-U}{k_B T})\{\exp[\frac{\overline{V}R}{k_B T}((\frac{\beta^C(T,B)}{\overline{V}})^{\frac{1}{2}}\overline{\alpha}\sqrt{\frac{T}{T_C - T}} - JB\frac{|v_{by}|}{v_T})]$$

$$-\exp[\frac{-\overline{V}R}{k_B T}((\frac{\beta^C(T,B)}{\overline{V}})^{\frac{1}{2}}\overline{\alpha}\sqrt{\frac{T}{T_C - T}} - JB\frac{|v_{by}|}{v_T})]\} \quad , \tag{17}$$

and



$$|v_{by}| = J\rho_{xx}/B \quad , \tag{18}$$

with $(\frac{\beta^C(T,B)}{\overline{V}})^{\frac{1}{2}}$ is the magnitude of the random collective pinning force per unit volume, $\overline{\alpha}$ is a proportional constant. It is clear that, from the above calculations, the directional-dependent potential barrier of the vortex bundle renormalized the Hall and longitudinal resistivities. By pondering the fact that the arguments in the exponential functions inside the curly bracket of Eqs. (16) and (17) are very small when the Lorentz force is closed to the random collective pinning force, we finally obtain the temperature-and field-dependent longitudinal and Hall resistivities as

$$\rho_{xx} = \frac{\overline{V}\sqrt{B\Phi_0}}{J\sqrt{\pi T}} \exp(\frac{-U}{k_B T})(\frac{2\overline{V}R}{k_B})[JB - (\frac{\beta^C(T,B)}{\overline{V}})^{\frac{1}{2}}] \quad , \tag{19}$$

$$\rho_{xy} = \frac{-\overline{V}\sqrt{B\Phi_0}}{J\sqrt{\pi T}} \exp(\frac{-U}{k_B T})(\frac{2\overline{V}R}{k_B})[(\frac{\beta^C(T,B)}{\overline{V}})^{\frac{1}{2}}\overline{\alpha}\sqrt{\frac{T}{T_C-T}} - JB\frac{|v_{by}|}{v_T}] \quad , \tag{20}$$

with $|v_{by}| = J\rho_{xx}/B$. Let us assume that the vortex system is initially in the region of the thermally activated motion of small vortex bundles, when the temperature of the system is kept at a constant value $T$, from our previous study[6], the value of $\frac{1}{B}(\frac{\beta^C(T,B)}{\overline{V}})^{\frac{1}{2}}$ increases with decreasing applied magnetic field, if it passes the value of $(\frac{|v_{by}|}{v_T})/(\overline{\alpha}\sqrt{\frac{T}{T_C-T}})$, then $\rho_{xy}$ changes sign from positive to negative with decreasing applied magnetic field. On the other hand, when the applied magnetic field is kept at a constant $B$, the value of $\frac{1}{B}(\frac{\beta^C(T,B)}{\overline{V}})^{\frac{1}{2}}$ increases and that of



$\bar{\alpha}\sqrt{\frac{T}{T_C - T}}$ decreases as temperature decreasing. Therefore, the term

$\frac{1}{B}(\frac{\beta^C(T,B)}{\bar{V}})^{\frac{1}{2}}\bar{\alpha}\sqrt{\frac{T}{T_C - T}}$ exists a maximum value at some temperature $T$, if this

maximum value is greater than $J\frac{|v_{by}|}{v_T}$, then $\rho_{xy}$ possesses the sign reversal

property. These results truly reflect the fact that the anomalous Hall effect is induced

by the competition between the Magnus force and the random collective pinning

force. It is interesting to note that $\rho_{xy}$ might possess the double sign reversal

property, the detailed discussion will be given in Sec. V. Moreover, since the value of

$\frac{1}{B}(\frac{\beta^C(T,B)}{\bar{V}})^{\frac{1}{2}}$ increases with decreasing temperature (applied magnetic field),

therefore, $\rho_{xx}$ decreases monotonically as temperature (applied magnetic field)

decreases. It is worthwhile pointing out that, from the above results, these anomalous

properties are independent of the mechanism for the superconductors as well as the

expression of $\bar{V}$, the volume for the vortex bundle. In other words, these anomalous

properties are universal for type-II conventional and high-$T_C$ superconductors as well

as for superconducting bulk materials and thin films, provided the conditions given

above are satisfied. The detailed calculations and discussion will be given in the

following sections.

As we have mentioned in the Introduction, when temperature (applied magnetic



field) decreases below $T_p(B_p)$, the quasiorder-disorder first-order phase transition temperature (magnetic field) of the vortex system, the region crosses over to that of thermally activated motion of large vortex bundles. In this region, the potential barrier $U$ generated by the randomly distributed strong pinning sites inside the bundle is large, in this case both the Hall and longitudinal resistivities decay to zero quickly with decreasing temperature (magnetic field).

## III. ANOMALOU HALL EFFECT FOR TYPE-II SUPERCONDUCTING BULK MATERIALS

For superconducting bulk materials, the volume $\bar{V}$ for the vortex bundle in Eqs. (19) and (20) is given as $\bar{V} = \pi R^2 L$, where $R$ ($L$) is the transverse (longitudinal) size of the vortex bundle. In this case, the longitudinal and Hall resistivities for type-II superconducting bulk materials therefore become,

$$\rho_{xx} = \frac{\bar{V}\sqrt{\Phi_0}\sqrt{B}}{J\sqrt{\pi}\sqrt{T}}\exp(\frac{-U}{k_B T})[\frac{2\pi R^3 L}{k_B}][JB - (\frac{\beta^C(T,B)}{\bar{V}})^{\frac{1}{2}}] \quad , \tag{21}$$

$$\rho_{xy} = \frac{-\bar{V}\sqrt{\Phi_0}\sqrt{B}}{J\sqrt{\pi}\sqrt{T}}\exp(\frac{-U}{k_B T})[\frac{2\pi R^3 L}{k_B}][(\frac{\beta^C(T,B)}{\bar{V}})^{\frac{1}{2}}\bar{\alpha}\sqrt{\frac{T}{T_C - T}} - JB\frac{|v_{by}|}{v_T}] \quad , \tag{22}$$

with $|v_{by}| = J\rho_{xx}/B$. As we have indicated in Sec. II, the above equations give rise to the phenomenon of anomalous Hall effect, namely, the value of $\rho_{xy}$ changes it sign from positive to negative with decreasing applied magnetic field (temperature). We will discuss the cases for constant temperature and constant applied magnetic field



separately as follows.

## A. Hall and longitudinal resistivities for constant temperature

Under the framework of our theory, it has been shown in Sec. II that $\rho_{xy}$ possesses the desired anomaly properties as applied magnetic field decreases, when the temperature of the system is kept at a constant value $T$. It is interesting to make numerical calculations of $\rho_{xy}$ and $\rho_{xx}$ for the thermally activated motion of small vortex bundles at $T = 91K$. The results are given in Table I. These numerical results do exhibit the expected properties of the Hall anomaly, and are in good agreement with the experimental data for $YBa_2Cu_3O_{7-\delta}$ high-$T_C$ bulk materials [21]. In obtaining the above results, the following data are employed,

$$R = 2\times10^{-8}m, \quad L = 10^{-6}m, \quad J = 10^6 \frac{A}{m^2}, \quad T_C = 92K, \quad v_T = 10^3 m/\sec, \quad \overline{v} = 10^{11} \sec^{-1},$$

$$\overline{\alpha} = 5.59\times10^{-5} T^{\frac{-1}{2}}, \quad \exp(\frac{-U}{k_B T}) = 2.07\times10^{-2} m, \quad (\frac{\beta^C(B=3.5)}{\overline{V}})^{\frac{1}{2}} = 3.4506\times10^6 N/m^3,$$

$$(\frac{\beta^C(3.03)}{\overline{V}})^{\frac{1}{2}} = 2.9849\times10^6 N/m^3, \quad (\frac{\beta^C(2.5)}{\overline{V}})^{\frac{1}{2}} = 2.4605\times10^6 N/m^3, \quad (\frac{\beta^C(2)}{\overline{V}})^{\frac{1}{2}} = 1.9668\times10^6 N/m^3,$$

$$(\frac{\beta^C(1.5)}{\overline{V}})^{\frac{1}{2}} = 1.4748\times10^6 N/m^3, \quad (\frac{\beta^C(1)}{\overline{V}})^{\frac{1}{2}} = 9.854\times10^5 N/m^3, \quad (\frac{\beta^C(0.75)}{\overline{V}})^{\frac{1}{2}} = 7.4042\times10^5 N/m^3,$$

$$(\frac{\beta^C(0.5)}{\overline{V}})^{\frac{1}{2}} = 4.915\times10^5 N/m^3.$$

From our [6] previous study, if the applied field decreases below 0.5 Tesla, the quasiorder-disorder first-order phase transition of the vortex system occurs. The region crosses over to that of the thermally activated motion of large vortex bundles. In this region, the potential barrier $U$ generated by the randomly distributed strong pinning sites inside the vortex bundle becomes large. Both the Hall and longitudinal



resistivities approach to zero quickly with decreasing applied magnetic field. These results are also in consistent with experimental data on $YBa_2Cu_3O_{7-\delta}$ high-$T_C$ bulk materials [21].

For clarity, our calculated results of $\rho_{xy}$ and $\rho_{xx}$ for thermally activated motion of small vortex bundles as functions of applied magnetic field at $T = 91K$ are plotted in Fig. 1(a) and Fig. 1 (b) respectively. These results do reflect the anomalous properties as expected and are in good agreement with the experimental data on $YBa_2Cu_3O_{7-\delta}$ high-$T_C$ bulk materials [21].

**B. Hall and longitudinal resistivities for constant applied magnetic field**

Based upon our theory, it has been shown in Sec. II that the Hall resistivity changes its sign from positive to negative as the temperature decreases for constant applied magnetic field. The numerical estimations of the Hall and longitudinal resistivities for thermally activated motion of small vortex bundles when the applied magnetic field is kept at a constant value $B = 2.24$ Tesla are given in Table II. These numerical results do manifest themselves the desired and unusual properties of the anomalous Hall effect, and are in consistent with the experimental data for $YBa_2Cu_3O_{7-\delta}$ high-$T_C$ bulk materials [21]. In arriving at the above results, the following data are employed approximately,

$R = 2\times10^{-8} m$, $L = 10^{-6} m$, $J = 10^6 \frac{A}{m^2}$, $T_C = 92K$, $v_T = 10^3 m/\sec$, $\bar{v} = 10^{11} \sec^{-1}$



$$\bar{\alpha} = 5.59 \times 10^{-5} T^{\frac{-1}{2}}, \quad \exp(\frac{-U}{k_B T}) = 2.07 \times 10^{-2}, \quad (\frac{\beta^C(T=91.6)}{\bar{V}})^{\frac{1}{2}} = 2.178 \times 10^6 \, N/m^3,$$

$$(\frac{\beta^C(91.3)}{\bar{V}})^{\frac{1}{2}} = 2.194 \times 10^6 \, N/m^3, \quad (\frac{\beta^C(91)}{\bar{V}})^{\frac{1}{2}} = 2.204 \times 10^6 \, N/m^3, \quad (\frac{\beta^C(90)}{\bar{V}})^{\frac{1}{2}} = 2.217 \times 10^6 \, N/m^3,$$

$$(\frac{\beta^C(89)}{\bar{V}})^{\frac{1}{2}} = 2.223 \times 10^6 \, N/m^3, \quad (\frac{\beta^C(88)}{\bar{V}})^{\frac{1}{2}} = 2.224 \times 10^6 \, N/m^3.$$

As we have stated from our [6] previous study, when the temperature decreases below $88K$, the quasiorder-disorder first-order phase transition of the vortex system takes place. The region crosses over to that of the thermally activated motion of large vortex bundles. In this region, the potential barrier $U$ generated by the randomly distributed strong pinning sites within the vortex bundle becomes large. Both the Hall and longitudinal resistivities reduce to zero quickly with decreasing temperature.

The above numerical results of $\rho_{xy}$ and $\rho_{xx}$ for the thermally activated motion of small vortex bundles as functions of temperature $T$ when $B = 2.24$ Tesla are plotted in Fig 2 (a) and Fig. 2(b) respectively. These results do reflect the anomalous properties as expected and are in good agreement with the experimental data on $YBa_2Cu_3O_{7-\delta}$ high-$T_C$ bulk materials [21].

## IV. Anomalous Hall effect for type-II superconducting films

We would like to discuss the anomalous Hall effect for type-II superconducting films. In this case, the volume $\bar{V}$ for the vortex bundle in Eqs. (19) and (20) is given by $\bar{V} = \pi R^2 d$, where $R$ is the transverse size of the vortex bundle, and $d$ is the



thickness of the film. Therefore, the Hall and longitudinal resistivities become,

$$\rho_{xx} = \frac{\overline{v}\sqrt{\Phi_0}\sqrt{B}}{J\sqrt{\pi}\sqrt{T}} \exp(\frac{-U}{k_B T})[\frac{2\pi R^3 d}{k_B}][JB - (\frac{\beta^C(T,B)}{\overline{V}})^{\frac{1}{2}}] \quad , \tag{23}$$

$$\rho_{xy} = \frac{-\overline{v}\sqrt{\Phi_0}\sqrt{B}}{J\sqrt{\pi}\sqrt{T}} \exp(\frac{-U}{k_B T}) [\frac{2\pi R^3 d}{k_B}][(\frac{\beta^C(T,B)}{\overline{V}})^{\frac{1}{2}}\overline{\alpha}\sqrt{\frac{T}{T_C - T}} - JB\frac{|v_{by}|}{v_T}] \quad , \tag{24}$$

with $|v_{by}| = J\rho_{xx}/B$. As we have discussed in Sec. II, the above equations give rise to the phenomenon of anomalous Hall effect for both the constant applied magnetic field and constant temperature. We will examine these cases separately as follows.

**A. Hall and longitudinal resistivities for constant temperature**

Under the framework of our theory, it has been shown in Sec. II that the Hall resistivity $\rho_{xy}$ changes its sign from positive to negative as the applied magnetic field decreases, when the temperature of the system is kept at a constant value $T$. We would like to make numerical calculations of $\rho_{xy}$ and $\rho_{xx}$ for the case of thermally activated motion of small vortex bundles. When temperature is kept at a constant value $T = 4.5K$, the Hall and longitudinal resistivities as functions of applied magnetic field in Tesla are given in Table III. These numerical results do possess the anomalous Hall properties as expected, and are in good agreement with the experiments values for $Mo_3Si$ conventional low-$T_C$ superconducting films [22]. In obtaining the above results, the following data are used approximately,

$R = 2\times 10^{-8} m$ , $d = 5\times 10^{-8} m$ , $J = 1.5\times 10^5 \frac{A}{m^2}$ , $T_C = 7.5 K$ , $v_T = 30 \, m/\sec$ ,

$\overline{\alpha} = 1.0449\times 10^{-3} T^{\frac{-1}{2}}$ , $\overline{v} = 10^{11} \sec^{-1}$ , $\exp(\frac{-U}{k_B T}) = 3.0899\times 10^{-4}$ , $(\frac{\beta^C(B=7.5)}{\overline{V}})^{\frac{1}{2}} = 4.5772\times 10^5 N/m^3$ ,



$(\frac{\beta^C(7.25)}{\bar{V}})^{\frac{1}{2}} = 4.4737 \times 10^5 \, N/m^3$ , $(\frac{\beta^C(7)}{\bar{V}})^{\frac{1}{2}} = 4.4324 \times 10^5 \, N/m^3$ , $(\frac{\beta^C(6.75)}{\bar{V}})^{\frac{1}{2}} = 4.3964 \times 10^5 \, N/m^3$ ,

$(\frac{\beta^C(6.5)}{\bar{V}})^{\frac{1}{2}} = 4.3483 \times 10^5 \, N/m^3$ , $(\frac{\beta^C(6.25)}{\bar{V}})^{\frac{1}{2}} = 4.272 \times 10^5 \, N/m^3$ , $(\frac{\beta^C(6)}{\bar{V}})^{\frac{1}{2}} = 4.0516 \times 10^5 \, N/m^3$ ,

$(\frac{\beta^C(5.75)}{\bar{V}})^{\frac{1}{2}} = 3.7418 \times 10^5 \, N/m^3$ , $(\frac{\beta^C(5.5)}{\bar{V}})^{\frac{1}{2}} = 3.335 \times 10^5 \, N/m^3$ .

When the applied magnetic field decreases below 5.5 Tesla, the quasiorder-disorder first-order phase transition of the vortex system takes place [6]. The region belongs to the thermally activated motion of large vortex bundles. In this case, the potential barrier $U$ generated by the randomly distributed strong pinning sites within the vortex bundle becomes large. Both $\rho_{xy}$ and $\rho_{xx}$ reduce to zero rapidly. These consequences are also in agreement with the experiments values for $Mo_3Si$ conventional low-$T_C$ superconducting films [22].

The calculated results of the Hall resistivity $\rho_{xy}$ and longitudinal resistivities $\rho_{xx}$ for the thermally activated motion of small vortex bundles as functions of applied magnetic field at $T = 4.5K$ are plotted in Fig. 3(a) and Fig. 3(b) respectively. These results do manifest themselves the anomalous properties as expected, and are in consistent with experimental data on $Mo_3Si$ conventional low-$T_C$ superconducting films [22].

B. **Hall and longitudinal resistivities for constant applied magnetic field**

Under the frame of our theory, it has been shown in Sec. II that the Hall resistivity $\rho_{xy}$ changes its sign from positive to negative as the temperature of the



system decreases, when the applied magnetic field is kept at a constant value. It is interesting to make numerical estimations of $\rho_{xy}$ and $\rho_{xx}$ for the thermally activated motion of small vortex bundles when $B = 2$ Tesla. The results are given in Table IV, and are in consistence with the experimental data on $YBa_2Cu_3O_{7-\delta}$ high-$T_C$ superconducting films [23]. In obtaining the above results, the following data are employed approximately,

$R = 2 \times 10^{-8} m$, $d = 5 \times 10^{-8} m$, $J = 10^6 \frac{A}{m^2}$, $T_C = 94 K$, $v_T = 10^2 m/\sec$, $\overline{\alpha} = 7.325 \times 10^{-4} T^{-\frac{1}{2}}$,

$\overline{v} = 10^{11} \sec^{-1}$, $\exp(\frac{-U}{k_B T}) = 2.01 \times 10^{-1}$, $(\frac{\beta^C(T=92)}{\overline{V}})^{\frac{1}{2}} = 1.9298 \times 10^6 N/m^3$,

$(\frac{\beta^C(91.5)}{\overline{V}})^{\frac{1}{2}} = 1.938 \times 10^6 N/m^3$, $(\frac{\beta^C(91)}{\overline{V}})^{\frac{1}{2}} = 1.9444 \times 10^6 N/m^3$, $(\frac{\beta^C(90.5)}{\overline{V}})^{\frac{1}{2}} = 1.9493 \times 10^6 N/m^3$,

$(\frac{\beta^C(90)}{\overline{V}})^{\frac{1}{2}} = 1.9531 \times 10^6 N/m^3$, $(\frac{\beta^C(89.5)}{\overline{V}})^{\frac{1}{2}} = 1.9556 \times 10^6 N/m^3$, $(\frac{\beta^C(89)}{\overline{V}})^{\frac{1}{2}} = 1.9575 \times 10^6 N/m^3$.

It is worthwhile mentioning that when the temperature decreases below $89 K$, the quasiorder-disorder first-order phase transition of the vortex system takes place [6], the region crosses over to that of thermally activated motion of large vortex bundles. In this region, the potential barrier $U$ generated by the randomly distributed strong pinning sites inside the vortex bundle becomes large. Therefore, both $\rho_{xy}$ and $\rho_{xx}$ decrease quickly to zero with deceasing temperature.

    The above calculated results of $\rho_{xy}$ and $\rho_{xx}$ for thermally activated motion of small vortex bundles as functions of temperature, when the applied magnetic field is kept at $B = 2$ Tesla, are plotted in Fig. 4(a) and Fig. 4(b) respectively. These results do display the anomalous properties as expected and are in good agreement with the



experimental data on $YBa_2Cu_3O_{7-\delta}$ high-$T_C$ superconducting thin films [23].

## V. REENTRY PHENOMENON FOR ANOMALOS HALL EFFECT

Under the framework of our theory, in this section we shall investigate the double sign reversal or the reentry phenomenon for the anomalous Hall effect. The crucial conditions for occurring this fascinating reentry phenomenon are that if the maximum value of $\frac{1}{B}(\frac{\beta^C(T,B)}{\bar{V}})^{\frac{1}{2}}\bar{\alpha}\sqrt{\frac{T}{T_C-T}}$ is greater than the value of $J\frac{|v_{by}|}{v_T}$, as we have discussed in Sec. II, and that if $T_P < T_R$, where $T_p$ is the quasiorder-disorder first-order phase transition [6] temperature, $T_R$ is the temperature for $\rho_{xy}$ crossing over back from negative to positive value, then the reentry phenomenon could happen. It is understood that $T_p$ decreases with increasing total random pinning force-the random collective pinning force plus the strong pinning force of the vortex system [24]. Hence, materials with larger total random pinning force, such as $YBa_2Cu_3O_{7-\delta}$ [25] or $Tl_2Ba_2Cu_2O_8$ [26], the reentry phenomenon could been observed.

It is interesting to make numerical estimations of the above situation for small vortex bundles. When the applied magnetic field $B$ is kept at a constant value $B=2$ Tesla, $\rho_{xy}$ and $\rho_{xx}$ as functions of temperature are given in Table V. The above numerical results do reflect the fascinating reentry phenomenon and are in agreement with the experimental data on $Tl_2Ba_2Cu_2O_8$ high-$T_C$ superconducting films [26]. In



obtaining the above results, the following data are used approximately,

$R = 2 \times 10^{-8} m$, $d = 10^{-6} m$, $J = 10^7 \frac{A}{m^2}$, $T_C = 104 K$, $v_T = 10^2 m/\sec$, $\bar{\alpha} = 1.12 \times 10^{-4} T^{-\frac{1}{2}}$,

$\bar{\nu} = 10^{11} \sec^{-1}$, $\exp(\frac{-U}{k_B T}) = 8.3199 \times 10^{-5}$, $(\frac{\beta^C(T=102)}{\bar{V}})^{\frac{1}{2}} = 1.8464 \times 10^7 N/m^3$,

$(\frac{\beta^C(100)}{\bar{V}})^{\frac{1}{2}} = 1.9031 \times 10^7 N/m^3$, $(\frac{\beta^C(98)}{\bar{V}})^{\frac{1}{2}} = 1.93267 \times 10^7 N/m^3$, $(\frac{\beta^C(96)}{\bar{V}})^{\frac{1}{2}} = 1.9512 \times 10^7 N/m^3$,

$(\frac{\beta^C(92)}{\bar{V}})^{\frac{1}{2}} = 1.955 \times 10^7 N/m^3$, $(\frac{\beta^C(88)}{\bar{V}})^{\frac{1}{2}} = 1.95618 \times 10^7 N/m^3$, $(\frac{\beta^C(84)}{\bar{V}})^{\frac{1}{2}} = 1.9588 \times 10^7 N/m^3$,

$(\frac{\beta^C(78)}{\bar{V}})^{\frac{1}{2}} = 1.961069 \times 10^7 N/m^3$, $(\frac{\beta^C(76)}{\bar{V}})^{\frac{1}{2}} = 1.9631 \times 10^7 N/m^3$.

It is worthwhile noting that when the temperature decreases below $76 K$, the quasiorder-disorder first-order phase transition of the vortex system takes place [6], the system crosses over to the region of thermally activated motion of large vortex bundles. In this case, the potential barrier $U$ generated by the randomly distributed strong pinning sites inside the vortex bundle becomes large. Both $\rho_{xy}$ and $\rho_{xx}$ decrease promptly to zero with decreasing temperature. These results are also in agreement with the experimental data on $Tl_2 Ba_2 Cu_2 O_8$ high-$T_C$ superconducting films [26].

The calculated results of $\rho_{xy}$ and $\rho_{xx}$ for thermally activated motion of small vortex bundles as functions of temperature, when the applied magnetic field is kept at $B = 2$ Tesla are plotted in Fig. 5(a) and Fig. 5 (b) respectively. These results do display the reentry properties as expected, and are in agreement with the experimental data on $Tl_2 Ba_2 Cu_2 O_8$ high-$T_C$ superconducting films [26].



## VI. Conclusion

We have developed a theory for the thermally activated motion of vortex bundles jumping over the directional-dependent potential barrier induced by the Magnus force, the random collective pinning force, and strong pinning force inside the vortex bundles for type-II superconductors. Under the framework of our theory, the anomalous Hall effect for conventional and high-$T_C$ superconducting bulk materials and thin films are investigated. It is shown that the Hall anomaly is universal for type-II conventional and high-$T_C$ superconductors as well as for superconducting bulk materials and thin films provided certain conditions are satisfied. We find that the directional-dependent potential barrier of the vortex bundles renormalizes the Hall and longitudinal resistivities and that the Hall anomaly is induced by the competition between the Magnus force and the random collective pinning force. We also find that the domain of anomalous Hall effect includes two regions: the region of thermally activated motion of the small vortex bundles and that of the large vortex bundles separated by the contour of the quasiorder-disorder first-order phase transition, or the peak effect [6] of the vortex system. Based on our theory, the crucial conditions for occurring double sign reversal are also investigated. Finally, the Hall and longitudinal resistivities are calculated for constant applied magnetic field as well as constant temperature, and the reentry phenomenon is also discussed. All the results



are in agreement with the experiments.

## ACKNOWLEDGMENTS

The authors would like to thank Professors Shiping Feng and E H Brandt for useful and constructive discussions.

TABLE

TABLE I. $\rho_{xx}$ and $\rho_{xy}$ as functions of applied magnetic field in Tesla for $YBa_2Cu_3O_{7-\delta}$ high-$T_C$ bulk materials at $T = 91\ K$.

Table II. $\rho_{xx}$ and $\rho_{xy}$ as functions of temperature for $YBa_2Cu_3O_{7-\delta}$ high-$T_C$ bulk materials at $B = 2.24$ Tesla.

Table III. $\rho_{xx}$ and $\rho_{xy}$ as functions of applied magnetic field in Tesla for $Mo_3Si$ conventional low-$T_C$ superconducting films at $T = 4.5\ K$.

Table IV. $\rho_{xx}$ and $\rho_{xy}$ as functions of temperature for $YBa_2Cu_3O_{7-\delta}$ high-$T_C$ superconducting films at $B = 2$ Tesla.

Table V. $\rho_{xx}$ and $\rho_{xy}$ as functions of temperature for $Tl_2Ba_2Cu_2O_8$ high-$T_C$ superconducting films at $B = 2$ Tesla.



FIGURES

FIG. 1. (a) The Hall resistivity $\rho_{xy}$ and (b) the longitudinal resistivity $\rho_{xx}$ of $YBa_2Cu_3O_{7-\delta}$ high-$T_C$ superconducting bulk materials for small vortex bundles as functions of magnetic field at $T = 91\,K$ with $T_C = 92K$. Both $\rho_{xy}$ and $\rho_{xx}$ reduce to zero quickly for the thermally activated motion of large vortex bundles when $B$ decreases below $0.5$ Tesla (not shown in the figure).

FIG. 2. (a) The Hall resistivity $\rho_{xy}$ and (b) the longitudinal resistivity $\rho_{xx}$ of $YBa_2Cu_3O_{7-\delta}$ high-$T_C$ superconducting bulk materials for small vortex bundles as functions of temperature $T$ at $B = 2.24$ Tesla with $T_C = 92K$. Both $\rho_{xy}$ and $\rho_{xx}$ approach to zero quickly for the thermally activated motion of large vortex bundles when $T$ decreases below $88\,K$ (not shown in the figure).

FIG. 3. (a) The Hall resistivity $\rho_{xy}$ and (b) the longitudinal resistivity $\rho_{xx}$ of $Mo_3Si$ conventional low-$T_C$ superconducting films for small vortex bundles as functions of magnetic field at $T = 4.5\,K$. Both $\rho_{xy}$ and $\rho_{xx}$ reduce to zero quickly for the thermally activated motion of large vortex bundles when $B$ decreases below $5.5$ Tesla (not shown in the figure).



FIG. 4. (a) $\rho_{xy}$ and (b) $\rho_{xx}$ of $YBa_2Cu_3O_{7-\delta}$ high-$T_C$ superconducting films for small vortex bundles as functions of temperature with $T_C = 94\ K$, when the applied magnetic field is kept at $B = 2$ Tesla. Both $\rho_{xy}$ and $\rho_{xx}$ approach to zero quickly for the thermally activated motion of large vortex bundles when $T$ decreases below $89\ K$ (not shown in the figure).

FIG. 5. (a) $\rho_{xy}$ and (b) $\rho_{xx}$ for the double sign reversal phenomena as functions of temperature for small vortex bundles when the applied magnetic field is kept at $B = 2$ Tesla. Both $\rho_{xy}$ and $\rho_{xx}$ approach to zero quickly for the thermally activated motion of large vortex bundles when $T$ decreases below $76\ K$ (not shown in the figure).



TABLES

Table I

| $B$ (T) | $\rho_{xy}$ ($\Omega$ m) | $\rho_{xx}$ ($\Omega$ m) |
|---|---|---|
| 3.5 | $1.2914 \times 10^{-9}$ | $1.8739 \times 10^{-6}$ |
| 3.03 | $1.7658 \times 10^{-15}$ | $1.5916 \times 10^{-6}$ |
| 2.5 | $-1.4627 \times 10^{-9}$ | $1.2663 \times 10^{-6}$ |
| 2.0 | $-2.7907 \times 10^{-9}$ | $9.5142 \times 10^{-7}$ |
| 1.5 | $-3.9996 \times 10^{-9}$ | $6.2539 \times 10^{-7}$ |
| 1.0 | $-4.6405 \times 10^{-9}$ | $2.9669 \times 10^{-7}$ |
| 0.75 | $-3.9801 \times 10^{-9}$ | $1.6825 \times 10^{-7}$ |
| 0.5 | $-2.0098 \times 10^{-9}$ | $1.226 \times 10^{-7}$ |

Table II

| $T$ (K) | $\rho_{xy}$ ($\Omega$ m) | $\rho_{xx}$ ($\Omega$ m) |
|---|---|---|
| 91.6 | $6.96 \times 10^{-10}$ | $1.878 \times 10^{-6}$ |
| 91.3 | $-1.82 \times 10^{-11}$ | $1.396 \times 10^{-6}$ |
| 91 | $-2.491 \times 10^{-9}$ | $1.094 \times 10^{-6}$ |
| 90 | $-4.199 \times 10^{-9}$ | $7.028 \times 10^{-7}$ |
| 89 | $-4.722 \times 10^{-9}$ | $5.22 \times 10^{-7}$ |
| 88 | $-3.081 \times 10^{-9}$ | $4.91 \times 10^{-7}$ |

Table III

| $B$ (T) | $\rho_{xy}$ ($\Omega$ m) | $\rho_{xx}$ ($\Omega$ m) |
|---|---|---|
| 7.5 | $4.3399 \times 10^{-11}$ | $8.2782 \times 10^{-7}$ |
| 7.25 | $1.5804 \times 10^{-11}$ | $7.8079 \times 10^{-7}$ |
| 7.0 | $-2.6037 \times 10^{-11}$ | $7.2721 \times 10^{-7}$ |
| 6.75 | $-6.7189 \times 10^{-11}$ | $6.742 \times 10^{-7}$ |
| 6.5 | $-1.023 \times 10^{-10}$ | $6.24 \times 10^{-7}$ |
| 6.25 | $-1.283 \times 10^{-10}$ | $5.78 \times 10^{-7}$ |
| 6.0 | $-1.1842 \times 10^{-10}$ | $5.492 \times 10^{-7}$ |
| 5.75 | $-8.8118 \times 10^{-11}$ | $5.3044 \times 10^{-7}$ |
| 5.5 | $-3.752 \times 10^{-11}$ | $5.2209 \times 10^{-7}$ |



Table IV

| T (K) | $\rho_{xy}$ (Ω m) | $\rho_{xx}$ (Ω m) |
|---|---|---|
| 92.0 | $2.0023\times10^{-9}$ | $9.7317\times10^{-7}$ |
| 91.5 | $4.9915\times10^{-10}$ | $8.6241\times10^{-7}$ |
| 91.0 | $-1.252\times10^{-9}$ | $7.7545\times10^{-7}$ |
| 90.5 | $-2.4984\times10^{-9}$ | $7.082\times10^{-7}$ |
| 90.0 | $-2.902\times10^{-9}$ | $6.5791\times10^{-7}$ |
| 89.5 | $-2.0005\times10^{-9}$ | $6.2461\times10^{-7}$ |
| 89.0 | $-7.5142\times10^{-10}$ | $5.9962\times10^{-7}$ |

Table V

| T (K) | $\rho_{xy}$ (Ω m) | $\rho_{xx}$ (Ω m) |
|---|---|---|
| 102 | $2.1346\times10^{-11}$ | $1.6728\times10^{-8}$ |
| 100 | $-3.69\times10^{-19}$ | $1.0657\times10^{-8}$ |
| 98 | $-1.4607\times10^{-11}$ | $7.4819\times10^{-9}$ |
| 96 | $-2.3518\times10^{-11}$ | $5.4754\times10^{-9}$ |
| 92 | $-1.0344\times10^{-11}$ | $5.1609\times10^{-9}$ |
| 88 | $5.9409\times10^{-19}$ | $5.1382\times10^{-9}$ |
| 84 | $5.46\times10^{-12}$ | $4.951\times10^{-9}$ |
| 78 | $1.3011\times10^{-11}$ | $4.8488\times10^{-9}$ |
| 76 | $1.2967\times10^{-11}$ | $4.65\times10^{-9}$ |



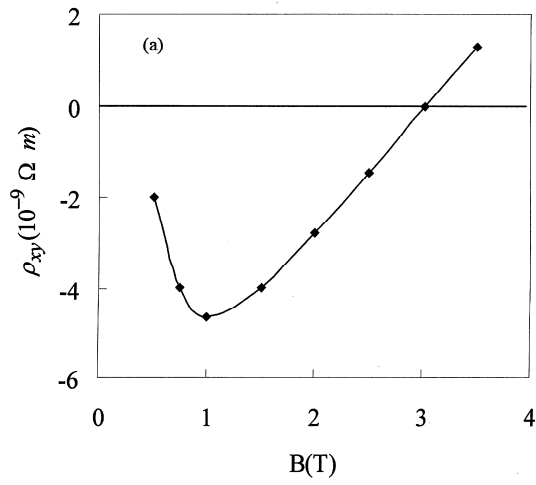

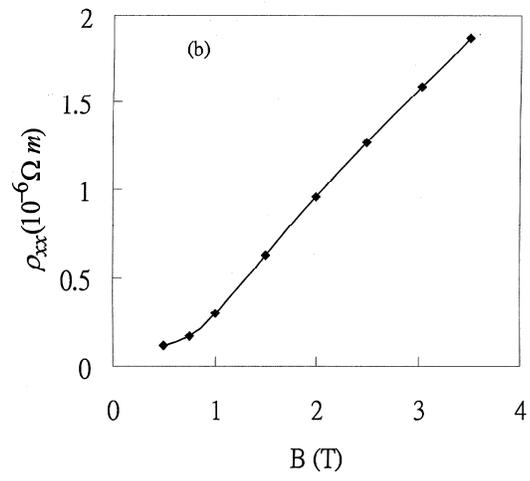

Fig. 1.



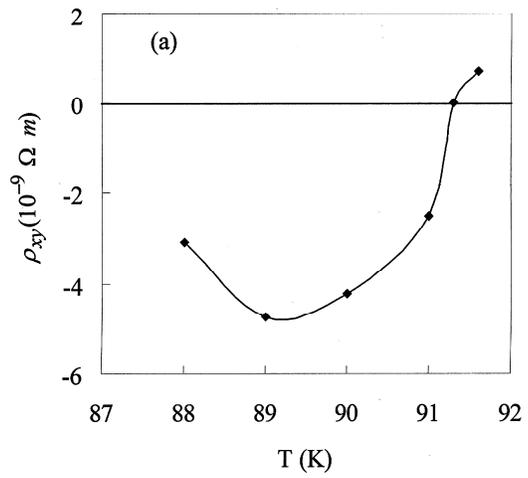

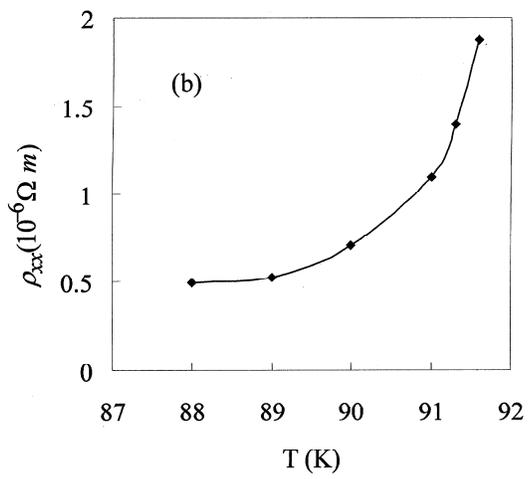

Fig. 2.



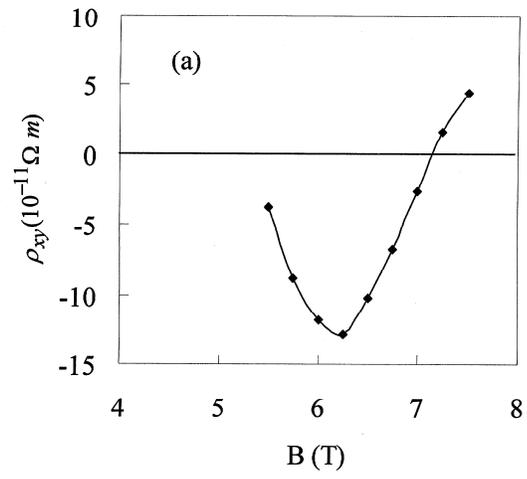

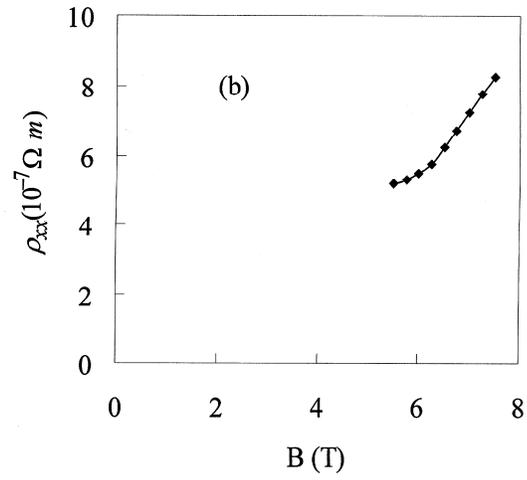

Fig. 3.



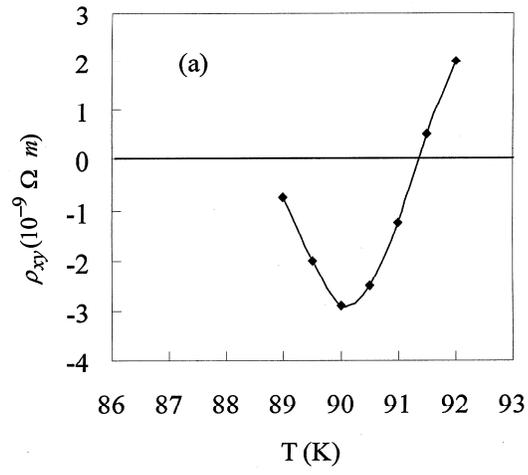

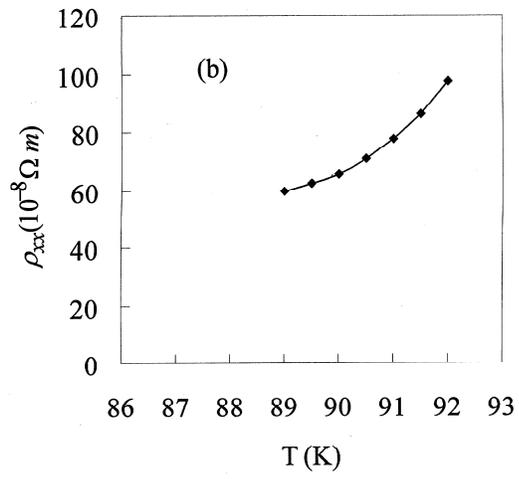

Fig. 4.



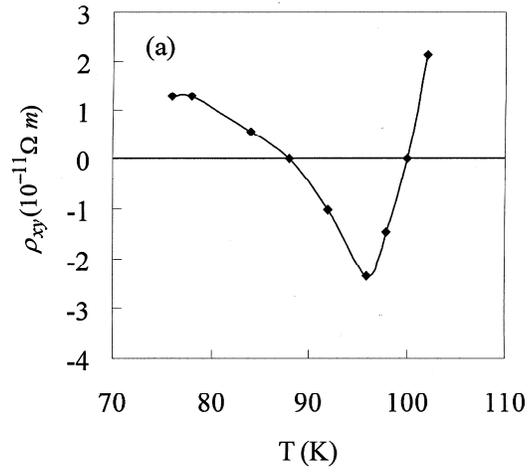
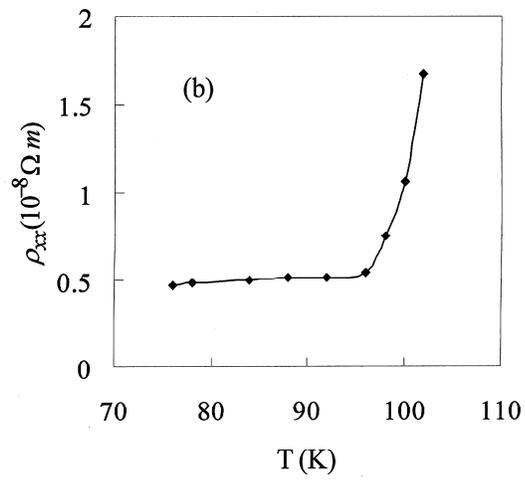

Fig. 5.